\documentstyle[epsfig,12pt]{article}
\def\bb#1{{\bf #1}}

\begin{document}

\begin{titlepage}
\begin{flushright}
\begin{minipage}{5cm}
\begin{flushleft}
\small
\baselineskip = 13pt YCTP-P6-00\\ TMUP-HEL-0011 \\ 
\end{flushleft}
\end{minipage}
\end{flushright}
\begin{center}
{\large\bf Aharonov-Bohm and Coulomb Scattering Near the Forward 
Direction}\footnote {Talk presented at Yale-TMU Symposium on 
Dynamics of Gauge Fields, Tokyo, Dec.\ 1999.}\\ \vskip 4ex Charles 
M.\ Sommerfield\footnote                  {\tt 
charles.sommerfield@yale.edu}\\ 
 Department of Physics, Yale 
University,\\ 
 New Haven, CT 06520-8120, USA\\
 \vskip2ex
Hisakazu Minakata\footnote{\tt minakata@phys.metro-u.ac.jp}   
\\ Department of Physics, Tokyo 
Metropolitan University,\\ Hachioji, Tokyo 192-0397, Japan\\{\em 
and}\\Research Center for Cosmic Neutrinos, Institute for Cosmic 
Ray Research, \\ University of Tokyo, Chiba 277-8582, Japan\\ 
\vskip4ex      June, 2000 
\end{center} 
\vfil
\begin{abstract}

The exact wave functions that describe scattering of a charged 
particle by a confined magnetic field (Aharonov-Bohm effect) and 
by a Coulomb field are analyzed.  It is well known that the usual 
procedure of finding asymptotic forms of these functions which 
admit a separation into a superposition of an incident plane wave 
and a circular or spherical scattered wave is problematic near the 
forward direction.  It thus appears to be impossible to express 
the conservation of probability by means of an optical theorem of 
the usual kind.  Both the total cross section and the forward 
scattering amplitude appear to be infinite.   To address these 
difficulties we find a new representation for the asymptotic form 
of the Aharonov-Bohm wave function that is valid for all angles.  
Rather than try to define a cross section at forward angles, 
however, we work instead with the probability current and find 
that it is quite well behaved. The same is true for Coulomb 
scattering.  We trace the usual difficulties to a nonuniformity of 
limits.     
\end{abstract}
\vfil
\end{titlepage}
\newpage
\section{Introduction}

Not only is 1999 the fiftieth anniversary of Tokyo Metropolitan 
University, it also marks forty years since the publication of the 
paper of Aharonov and Bohm \cite{ABeffect}  in which it was shown 
that there are measurable effects that can be attributed  directly 
to the electromagnetic vector potential and only non-locally to 
the magnetic field itself.  Of course in non-Abelian gauge 
theories, which are the subject of this conference, the direct 
importance of the potential does not raise any eyebrows, but it 
did so with respect to electromagnetism at the time.  The effect 
was soon confirmed experimentally by Chambers \cite{Chambers} and 
subsequently by many others \cite{Tonomura}.  In connection with 
anniversaries it may be interesting to note that in 1949 (again, 
fifty years ago) Ehrenberg and Siday published a paper 
\cite{Ehrenberg} in which the ray optics of an electron moving 
near a magnetic field were analyzed.  It is possible to view this 
work  as a precursor to that of Aharonov and Bohm. 

This paper by no means represents a comprehensive review. The 
literature of the subject is enormous.  The reader can use the 
references we have provided to help trace its detailed history. We 
begin in Sec.\ 2 with a brief review of the Aharonov-Bohm effect.  
What is involved is the effect of the magnetic field on a 
charged-particle wave function if the particle is forbidden from 
entering the region where the field strength is  not zero.  We 
give a heuristic derivation of the influence of the field on the 
interference pattern of a double-slit experiment.  Aharonov and 
Bohm showed that if the magnetic field is restricted to an 
infinite line and if the wave function of the particle is chosen 
to  vanish on that line, then the problem is exactly soluble.  
Almost all work on the subject starts from their solution. The 
wave function is given as an infinite sum of partial waves.  The 
bulk of our presentation consists of an analysis of this sum. 

In Sec.\ 3 we apply the usual method of partial wave analysis to 
calculate the phase shifts.  This also leads to a formal 
derivation of the  optical theorem.  However, since the phase 
shifts are constant for large partial waves the convergence of the 
procedure is suspect.   In Sec.\ 4 we review other attempts to 
find the asymptotic form of the Aharonov-Bohm function and to 
reconcile the results with the optical theorem.  Sec.\ 5 is 
devoted to a new presentation of the Aharonov-Bohm function. It 
has an asymptotic form which is valid at all angles. In Sec.\ 6 we 
present graphical results of the probability current distribution 
from which one usually calculates cross sections.  The behavior 
near forward angles is quite finite and closely follows the 
double-slit interference pattern mentioned above. 

In Sec.\ 7 we show that Coulomb scattering shares many of the same 
features of Aharonov-Bohm scattering.  Again we provide graphical 
results of the probability current as derived from the exact 
Coulomb wave functions. 

\section{The Aharonov-Bohm effect and a simple  model}

Consider a charged particle in the presence of a magnetic field 
${\bb B}({\bb r})$ which is described by a vector potential ${\bb 
A}({\bb r})$ such that ${\bf B}=\nabla\times{\bf A}$. 
 If in the absence of the magnetic field the system 
is described by a stationary-state wave function $\psi_0(\bb r)$, 
then in the presence of the field it will be described by the 
path-dependent multiple-valued wave function 
\begin{equation}
 \psi(\bb r)=\psi_0(\bb r)\exp\left[i{q\over\hbar c}
 \int_{\bb r_0}^{\bb r} \bb A(\bb r')\cdot d\bb r'\right]=\psi_0(\bb r)
 e^{i\chi(\bb r,P)}. 
\end{equation}   
Here $q$ is the charge of the particle and $P$ symbolizes the path 
followed by the integral.  

\begin{figure}[h]
\epsfxsize=17.25pc \centerline{\epsfbox{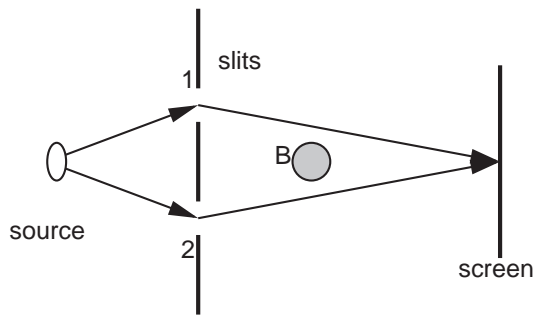}} \caption{The 
Aharonov-Bohm double slit experiment.} \label{fig.1} 
\end{figure}

If one arranges a double slit experiment, as in Fig.\ 1, so that 
the part of the beam that goes through 
 slit 1 passes entirely above the magnetic field while the part 
that goes through slit 2 passes entirely below the field then 
using a ray-optics approximation one expects that at the 
particular position that would be the central point of the 
interference pattern in the absence of the field,  the wave 
function will now be 
\begin{eqnarray}
 \psi(\bb r)&=&\hbox{$1\over2$}\psi_0(\bb r)[e^{i\chi(\bb r,P_1)}
 +e^{i\chi(\bb r,P_2)}]\nonumber\\
 &=&\hbox{$1\over2$}\psi_0(\bb r)e^{i\chi(\bb r,P_1)}
 \left[1+\exp\left( i{q\Phi\over\hbar c}\right)\right],
 \end{eqnarray}   
where $\Phi$ is the flux of the magnetic field through the area 
bounded by the two paths.  Thus the interference pattern will 
change with the magnetic field, even if the electron wave function 
is zero in the region to which the field is confined.  This 
contradicts the classical notion that the vector potential is 
basically unphysical and that a charged particle can  feel a force 
only at points where the magnetic field itself is nonzero. 

The double slit geometry is not essential for the Aharonov-Bohm 
effect.  Aharonov and Bohm treated an idealization of the 
situation without the slits that could be solved analytically.  
They restricted the magnetic field to be confined to a line of 
zero width oriented along the $z$ direction.  That is they took 
\begin{equation}
  \bb B(\bb r)=\hat\bb z\,\Phi\,\delta(x)\delta(y).\label{simple}
\end{equation} 
In this case, using plane polar coordinates $r$ and $\theta$ in 
the $x$-$y$ plane, one may take the vector potential to be 
\begin{equation}
  \bb A(r,\theta)={\Phi\over2\pi 
   r}\hat{\mbox{\boldmath $\theta$}}={\Phi\over2\pi}\nabla\theta.
\end{equation}
  The many-valued
wave function in the presence of the field would be simply 
 \begin{equation}
 \psi(r,\theta)=\psi_0(r,\theta)\exp\left[i{q\over\hbar c}{\Phi\over2\pi}\theta\right]=
 \psi_0(\theta)e^{i\alpha\theta}.\label{multi}
 \end{equation} 
We have defined $\alpha$ to measure the strength of the flux.  It 
is convenient for later developments to take the greatest integer 
in $\alpha$ to be $N$ and to write 
 \begin{equation}
 \alpha={q\over\hbar 
 c}{\Phi\over2\pi}=[\alpha]+\hbox{$1\over2$}+\gamma=N+\hbox{$1\over2$}+\gamma,
 \end{equation}
with $-{1\over2}<\gamma<{1\over2}$.  

The Aharonov-Bohm scattering problem is to find a single-valued 
solution of Schr\"odinger's equation that corresponds as closely 
as possible to an incident plane wave moving to the left along the 
$x$ axis from the right of the magnetic field in Fig.\ 1 (with the 
source, screen and slits eliminated) subject to the boundary 
condition that the wave function vanishes wherever the field is 
nonzero.   For the field of Eq.\ (\ref{simple}) we must take the 
wave function to vanish at the origin. Ignoring   the $z$ 
coordinate, the Schr\"odinger equation is 
\begin{equation}
\left[\left(\nabla-i{q\over\hbar c}\bb A(\bb 
r)\right)^2+k^2\right]\psi(r,\theta) =0, 
\end{equation}
which simplifies to 
\begin{equation}
\left[{\partial^2\over \partial r^2}+{1\over r}{\partial\over 
\partial r} +{1\over 
r^2}\left({\partial\over\partial\theta}-i\alpha\right)^2 
+k^2\right]\psi(r,\theta)=0.\label{sch} 
\end{equation}
The wave function in Eq.\ (\ref{multi}), which in this case is                                                                                                   
$\psi(r,\theta)=e^{-ikr\cos\theta}e^{i\alpha\theta}$,                                                                                                            
is clearly not acceptable despite the fact it solves Eq.\ 
(\ref{sch}),                          
 since it does not vanish at the origin and is not single-valued.      
A single-valued solution can be found by a partial wave expansion 
and becomes a superposition over all $l$ of solutions 
$J_{l-\alpha}(kr)e^{il\theta}$ that are eigenfunctions of the $z$ 
component of angular momentum. The wave function of Aharonov and 
Bohm  is of this form and does reduce to the simple plane wave 
when $\alpha=0$. It is 
\begin{equation}
\psi_{\rm AB}(r,\theta)=\sum_{l=-\infty}^{\infty} e^{-i\pi 
|l-\alpha|/2}J_{|l-\alpha|}(kr)e^{i l\theta}.\label{pw} 
\end{equation}   
Using a concept he calls ``whirling waves,'' Berry \cite{Berry} 
has given a beautiful derivation of this result that reconciles 
the requirement of single-valuedness with the  notion that the 
effect of the magnetic field 
 can be given by a simple multiplicative phase factor.
   
\section{Two-dimensional partial-wave expansion}

The asymptotic form for large $r$ of a wave function such as that 
given in Eq.\ (\ref{pw}) is expected to have the form 
\begin{equation}
\psi_{\rm AB}(r,\theta)\rightarrow 
e^{-ikr\cos\theta}+e^{i{\pi\over4}}f(\theta){e^{ikr}\over\sqrt{r}}, 
\end{equation}
where the scattering amplitude $f(\theta)$ would have the 
partial-wave expansion 
\begin{equation}
f(\theta)=\sqrt{2\over\pi k}\sum_{l=-\infty}^{\infty}(-1)^l 
e^{i\delta_l}\sin\delta_l\,e^{il\theta}. 
\end{equation}
An examination of Eq.\ (\ref{pw}) shows  the phase shifts to be 
$\delta_l= {\pi\over2}(l-|l-\alpha|)$ so that 
\begin{equation}
f(\theta)=\sqrt{2\over\pi k}\sin{\pi\over2}\alpha 
\left[e^{i{\pi\over2}\alpha}\sum_{l=N+1}^\infty  
(-1)^le^{il\theta}e^{-2\epsilon l} 
-e^{-i{\pi\over2}\alpha}\sum_{l=-\infty}^N (-1)^l 
e^{il\theta}e^{2\epsilon l}\right], 
\end{equation}
where convergence factors involving the small positive quantity 
$\epsilon$ have been inserted to insure convergence.  The sums are 
easily done and the result is 
\begin{equation}
f(\theta)={1\over\sqrt{2\pi 
k}}(-1)^Ne^{i(N+{1\over2})\theta}\left[{-\sin\pi\alpha 
\cos{\theta\over2} 
+(1-\cos\pi\alpha)\epsilon\sin{\theta\over2}\over\cos^2{\theta\over2}+\epsilon^2}\right]. 
\end{equation}
Letting $\epsilon\rightarrow0$ we find 
\begin{equation}
f(\theta)=-{\sin\pi\alpha\over\sqrt{2\pi 
k}}(-1)^Ne^{i(N+{1\over2})\theta}{1\over\cos{\theta\over2}}+i\pi{1-\cos\pi\alpha\over\sqrt{2\pi 
k}}\delta(\cos\hbox{${\theta\over2}$}), 
\end{equation}
which is a result obtained by Ruijsenaars \cite{Ruijsenaars}.  
Criticism of this formal method of summing the partial-wave 
expansion has been given by Hagen  \cite{Hagen}.  Arai and 
Minakata \cite{Arai} have analyzed this expression for $f(\theta)$ 
in connection with the optical theorem in great detail without a 
completely satisfactory conclusion. On the other hand Sakoda and 
Omote have shown that the S-matrix is indeed unitary  
\cite{Sakoda}.  

\section{Integral representations of the Aharonov-Bohm wave 
function} 

Another method that makes sense of the sum in Eq.\ (\ref{pw}) is 
to use a modified version of Sommerfeld's representation of the 
Bessel function 
 \begin{equation} 
 J_\nu(z)={1\over2\pi} e^{i\nu{\pi\over2}}\int_{\cal C}
 dt\; e^{-iz\cos t}e^{i\nu t},\label{C}
 \end{equation} 
where ${\cal C}$ goes from $-\pi+\eta+i\infty$ to 
$\pi+\eta+i\infty$ with $\eta$ a positive infinitesimal as 
illustrated in Fig.\ 2. 

\begin{figure}[t]
\epsfysize=1.2in \centerline{\epsfbox{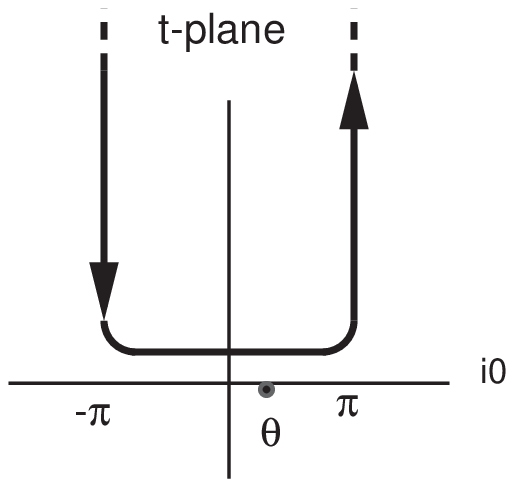}} \caption{The 
contour ${\cal C}$ of Eq.\ (\ref{C}).} \label{fig.2} 
\end{figure}

\noindent When this representation is used in Eq.\ (\ref{pw}) one 
can perform  the sum over $l$ under the integral sign to obtain 
\begin{equation}
\psi_{\rm AB}(r,\theta)={i\over 4\pi}e^{(N+{1\over2})\theta} 
\int_{\cal C}dt\;e^{-ikr\cos t}\left[{e^{i\gamma 
t}\over\sin{1\over2}(t-\theta)}+{e^{-i\gamma 
t}\over\sin{1\over2}(t+\theta)}\right].\label{contour} 
 \end{equation}

One way to proceed is to  note that the substitution 
$t\rightarrow-t$ in the second term of the integral in Eq.\ 
(\ref{contour}) leads to an expression identical with the first 
term but  integrated now along a contour that is ${\cal C}$ 
reflected in the origin.  The contours may be distorted, as shown 
by comparison of the left- and right-hand illustrations of Fig.\  
3, into two straight-line paths and a clockwise contour around  
the pole at $t=\theta$.  

\begin{figure}[h]
\epsfysize=1.8in\centerline{\epsfbox{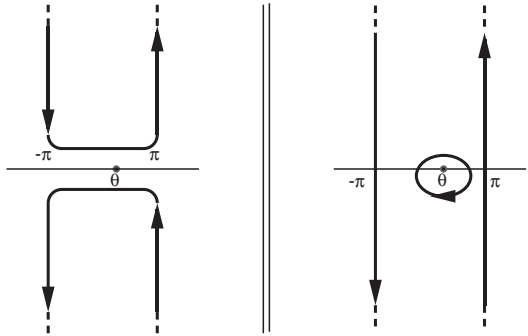}} 
\caption{Integration contours for Eq.\ (\ref{contour}).} 
\label{fig.3} 
\end{figure}
The contour around the pole gives rise to a modified incident wave 
of the form $e^{-ikr\cos\theta}e^{i\alpha\theta}$ which is not 
single-valued.  The other terms may be analyzed by 
steepest-descent methods and give rise to a scattered wave that is 
valid except for $\theta$ near $\pm\pi$.  The result is 
 \begin{equation}
 \psi_{\rm AB}(r,\theta)\rightarrow e^{-ikr\cos\theta}e^{i\alpha\theta}
-e^{i{\pi\over4}}{e^{ikr}\over\sqrt{r}}{1\over\sqrt{2\pi k}}(-1)^N 
\sin\alpha\pi 
{e^{i(N+{1\over2})\theta}\over\cos{1\over2}\theta}.\label{oldasymp} 
\end{equation}
Although the scattered wave is  not single-valued,  the entire 
wave function is.  It remains unclear from this approach, however,  
how to handle  scattering near the forward direction. 

\section{A new integral representation}

We go back to the original contour ${\cal C}$ and write 

\begin{equation}
\psi_{\rm 
AB}(r,\theta)={i\over4\pi}e^{i(N+{1\over2})\theta}e^{-ikr\cos\theta} 
A(kr,\theta), 
 \end{equation}
which, by comparison with Eq.\ (\ref{contour}) defines 
$A(kr,\theta)$. By differentiating $A(kr,\theta)$ with respect to 
$kr$  we can eliminate the singularities.  We find, after some 
manipulation of the resulting Bessel functions 
 \begin{eqnarray}
 \lefteqn{{\partial\over\partial (kr)}A(kr,\theta)=-2\pi
i\cos\gamma\pi\, e^{ikr\cos\theta}}\hspace{.5in}\nonumber\\ 
&&\times\left[e^{i{1\over2}\theta}e^{i({1\over2}+\gamma){\pi\over2}} 
H^{(1)}_{{1\over2}+\gamma}(kr) 
+e^{-i{1\over2}\theta}e^{i({1\over2}-\gamma){\pi\over2}} 
H^{(1)}_{{1\over2}-\gamma}(kr)\right], 
 \end{eqnarray} 
in which Hankel functions appear.  Since the wave function must 
vanish for $r=0$ we integrate  this result  from 0 to $kr$ and 
obtain the following exact expression for the Aharonov-Bohm wave 
function. 
 \begin{eqnarray}
\lefteqn{ \psi_{\rm AB}(r,\theta)=\hbox{$1\over2$}e^{-i{\pi\over 
4}}e^{i(N+{1\over2})\theta}\cos\gamma\pi\, 
e^{-ikr\cos\theta}}\nonumber\\ &&\times\int_0^{kr} 
dz\;e^{iz\cos\theta} 
\left[e^{i{1\over2}\theta}e^{i\gamma{\pi\over2}}H^{(1)}_{{1\over2}+\gamma}(z) 
+e^{-i{1\over2}\theta}e^{-i\gamma{\pi\over2}}H^{(1)}_{{1\over2}-\gamma}(z)\right]. 
\label{exact}  
\end{eqnarray}
It should be noted that this expression is valid only for 
$-\pi\le\theta\le\pi$.  For other values of $\theta$ one should 
replace $\theta$  by $[(\theta+\pi)\hbox{mod}(2\pi)]-\pi$. 
 
The asymptotic form of $\psi_{\rm AB}(r,\theta)$ for large $r$ and 
any value of $\theta$ may be found by writing the integrals from 0 
to $kr$ in Eq.\ (\ref{exact}) as the difference of integrals from 
0 to $\infty$ and from $kr$ to $\infty$.  For large $kr$ we may 
use the large-argument asymptotic form of the Hankel functions to 
approximate the second integral (no matter what the value of 
$\theta$) while the first integral can be done exactly 
\cite{Abramowitz}.  The result is 
\begin{equation}  
\psi_{\rm AB}(r,\theta)\rightarrow e^{i(N+{1\over2})\theta} 
e^{-ikr\cos\theta} \left\{e^{i\gamma\theta}-\cos\gamma\pi\left[1 
-\hbox{erf}\left(e^{-i{\pi\over4}}\sqrt{2kr}\cos\hbox{${1\over2}$}\theta)\right)\right]\right\}. 
\label{asymptotic} 
\end{equation}

Stelitano \cite{Stelitano} has obtained a similar expression by 
differentiating the partial-wave sum directly with respect to 
$kr$.  But he treats the $l=0$ term separately which means that 
his asymptotic expression is not as compact as Eq.\ 
(\ref{asymptotic}) nor is it strictly correct for backward 
scattering angles. Alvarez \cite{Alvarez} has also given an 
asymptotic form for large $r$ as an infinite series that is 
applicable for any $\theta$.

For $\sqrt{kr}\cos{1\over2}\theta\gg1$, (non-forward angles), Eq.\ 
(\ref{asymptotic}) reduces to Eq.\ ({\ref{oldasymp}), in agreement 
with the other integration method, while for $\theta=\pi$ and 
large $kr$ we get 
 $\psi(r,\pi)\rightarrow\cos\alpha\pi\, e^{ikr}$,
 which is perfectly finite and which exhibits the Aharonov-Bohm interference.

In view of these results it is to be stressed once again, as has 
been done by many previous workers on this problem, that near the 
forward direction the total wave function does not admit  a 
separation into incident and scattered parts.  Hence the usual 
statement of the optical theorem as an expression of unitarity 
just doesn't make sense. 

\section{Properties of the probability current}

From the time-independent point of view the usual optical theorem 
reflects the conservation of the probability current, the 
integrated form of which is $\int_{-\pi}^{\pi} d\theta\; \hat{\bb 
r}\cdot\bb j(\bb r)=0$, for any fixed $r$. The probability current 
density is given by 
 \begin{equation}
 \bb j(\bb r)={\hbar\over m}\hbox{Im}[\psi^*(\bb r)\nabla\psi(\bb 
 r)].
 \end{equation}
It usually has three terms---coming  from the incident wave, the 
scattered wave and the interference between them.  In a 
wave-packet treatment the latter has significance only near the 
forward direction.  Because it is proportional to $1/r$ the 
scattered term will be quite small compared to the incident term.  
And even at forward angles it will be overwhelmed by the incident 
term.  

\begin{figure}[b] \centerline{\epsfbox{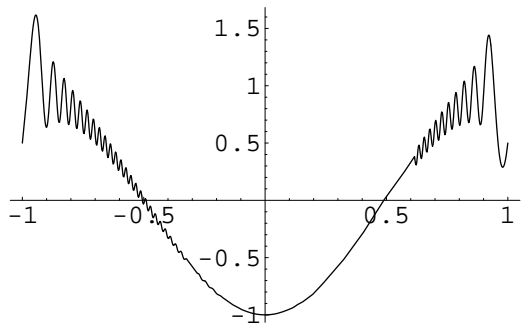}} \caption{A plot 
of  $\hat\bb r\cdot \bb j(\bb r)$ against ${\theta/\pi}$ for 
$kr=100$ and $\gamma=.25$, normalized to $-1$ at $\theta=0$.} 
\label{fig.4} 
\end{figure}

The large $kr$ approximation given by Eq.\ (\ref{asymptotic}) 
greatly simplifies the numerical calculation of  $\hat\bb 
r\cdot\bb j(\bb r)$.  The results are plotted in Fig.\ 4 for 
$\gamma=.25$ and $kr=100$, while in Fig.\ 5 we show the naively 
defined cross section for the same parameters.   What we see in 
the Aharonov-Bohm case, where we cannot make the separation into 
three terms, is that effectively the forward scattering becomes 
significant enough to completely modify the behavior in the 
forward direction, while the probability current remains perfectly 
well-behaved and continuous there.      

\begin{figure}[h]
\centerline{\epsfbox{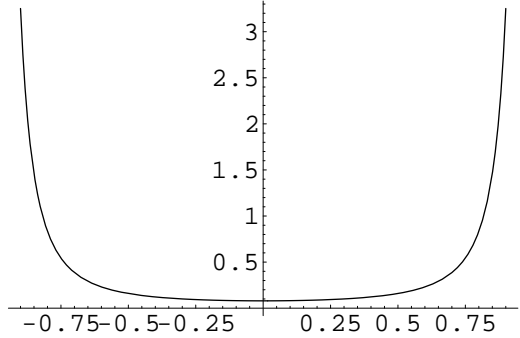}} \caption{A plot of the usual 
differential cross section in units of $1/k$ against $\theta/\pi$ 
for $kr=100$ and $\gamma=.25$.} \label{fig.5} 
\end{figure}

It is clear from this analysis that  the limits of 
$\theta\rightarrow\pi$ and $r\rightarrow\infty$ do not commute. It 
is our contention that the former limit is what makes sense 
physically for any finitely contained experiment.

\section{Coulomb scattering}

A similar nonuniformity of limits appears in the consideration of 
Coulomb scattering.   The exact wave function, written in terms of 
a confluent hypergeometric function is 
\begin{equation}
 \psi_{\rm Coul}(\bb r)=e^{ikr\cos\theta}\;{}_1F_1(-i\eta;1;ikr[1-\cos\theta]),
\end{equation}
with  Coulomb parameter $\eta=mZe^2/(\hbar k)$.  For large 
$kr(1-\cos\theta)$ this has the  asymptotic form 
 \begin{equation}
\psi_{\rm Coul}(\bb r)\sim e^{i\eta\log kr(1-\cos\theta)} 
 \left[e^{ikr\cos\theta}-{e^{ikr}\over kr}{e^{-2i\eta\log kr}\over
 (1-\cos\theta)^{1+2i\eta}}{\Gamma(1+i\eta)\over\Gamma(1-i\eta)}\eta\right],
 \end{equation}
 which looks like a plane wave incident along the $z$ axis plus a spherical wave, 
 but with Coulomb distortion.  The ``$r$-dependent'' scattering amplitude 
 extracted from this is  
\begin{equation}
 f(\theta)=-{\eta\over 
 k}{\Gamma(1+i\eta)\over\Gamma(1-i\eta)}{e^{-2i\eta\log kr}\over
 (1-\cos\theta)^{1+2i\eta}}
 \end{equation}
 and the differential cross section is
 \begin{equation}
 {d\sigma\over d\Omega}={\eta^2\over k^2(1-\cos\theta)^2}.
 \end{equation}
 When extrapolated to forward angles these are both infinite.
 
 On the other hand, the exact wave function and the probability 
 current are quite finite at $\theta=0$.  In fact the probability 
 current becomes very small compared to its maximum value at 
 near-forward angles.  This is illustrated in Fig.\  6 with $\eta=1.5$.

\begin{figure}[h]
\centerline{\epsfbox{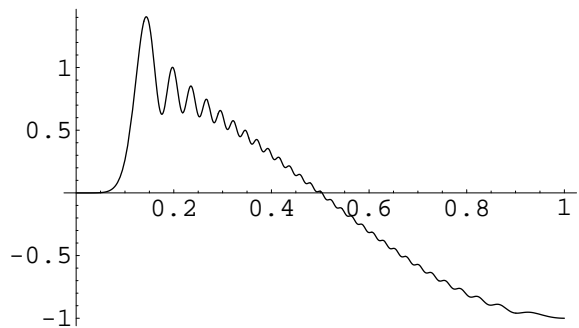}} \caption{Radial component of the 
probability current for Coulomb scattering, plotted against 
$\theta/\pi$ and normalized to $-1$ 
 at backward angles.  (The forward 
angle is now at $\theta=0$).  The parameters are $kr=100$ and 
$\eta=1.5$.} \label{fig.6} 
\end{figure}

\section*{Acknowledgments}
This work was supported in part by the U.S. Department of Energy 
under contract No.\ DE-FG-02-92ER-10701 and  by Grant-in-Aid for 
Scientific Research No.\ 09045036 under the International 
Scientific Research Program, Inter-University Cooperative 
Research, Japanese Ministry of Education, Science, Sports and 
Culture.

\end{document}